\documentclass[%
 reprint,
 superscriptaddress,
 amsmath,amssymb,
 aps,
]{revtex4-2}

\usepackage{graphicx}% Include figure files
\usepackage{dcolumn}% Align table columns on decimal point

\usepackage[usenames]{color} %used for font color

\usepackage{bm}% bold math
\usepackage{hyperref}% add hypertext capabilities
\begin{document}

\preprint{APS/123-QED}

\title{Keep the bees off the trees: The particular vulnerability of species \\ in the periphery of mutualistic networks to shock perturbations}
% Force line breaks with \\

\author{Lukas Halekotte}
\email{lukas.halekotte@uol.de}
\affiliation{Institute for Chemistry and Biology of the Marine Environment, Carl von Ossietzky University Oldenburg, 26129 Oldenburg, Germany}
 \affiliation{German Aerospace Center (DLR), Institute for the Protection of \\Terrestrial Infrastructures, 53757 Sankt Augustin, Germany}

\author{Anna Vanselow}
\affiliation{Institute for Chemistry and Biology of the Marine Environment, Carl von Ossietzky University Oldenburg, 26129 Oldenburg, Germany}

\author{Ulrike Feudel}
 \affiliation{Institute for Chemistry and Biology of the Marine Environment, Carl von Ossietzky University Oldenburg, 26129 Oldenburg, Germany}

\date{March 4, 2024}

\begin{abstract} 
We study the phenomenon of multistability in mutualistic networks of plants and pollinators, where one desired state in which all species coexist competes with multiple states in which some species are gone extinct. In this setting, we examine the relation between the endangerment of pollinator species and their position within the mutualistic network. To this end, we compare endangerment rankings which are derived from the species' probabilities of going extinct due to random shock perturbations with rankings obtained from different network theoretic centrality metrics. We find that a pollinator's endangerment is strongly linked to its degree of mutualistic specialization and its position within the core-periphery structure of its mutualistic network, with the most endangered species being specialists in the outer periphery. Since particularly well established instances of such peripheral areas are tree-shaped structures which stem from links between nodes/species in the outermost shell of the network, we summarized our findings in the admittedly ambiguous slogan \textit{keep the bees off the trees}. Finally, we challenge the generality of our findings by testing whether the title of this work still applies when being located in the outer periphery allows pollinators to avoid competitive pressure.
\end{abstract}


\maketitle

\section{Introduction} \label{introduction}
\begin{figure}[htb]
% \centering
\includegraphics[width=1.0\linewidth]{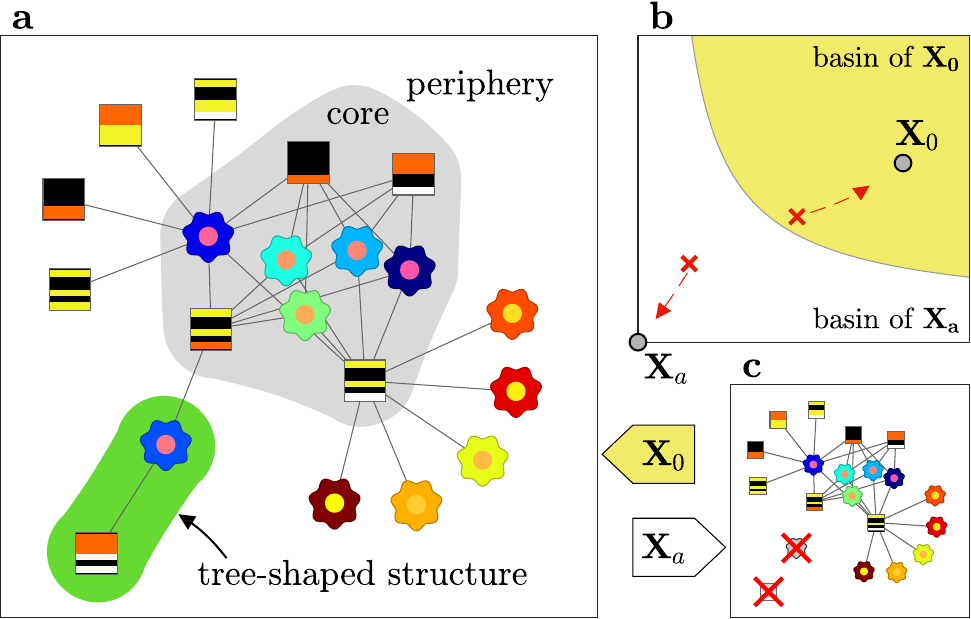}
\caption{\textbf{Core-periphery structure and multistability in a plant-pollinator network.} (a) Exemplary plant-pollinator network exhibiting a core-periphery structure with a single tree-shaped part (highlighted in green). The network is bipartite with links existing only between plant (flowers) and pollinator (squares) species. (b) Exemplary phase space with two basins of attraction, one corresponding to the desired state $\mathbf{X}_0$ (yellow) in which all species coexist and one to an alternative undesired state $\mathbf{X}_a$ (white) in which some species are extinct. The two initial conditions (red crosses) depict a non-fatal and a fatal perturbation leading to the two opposing attractive states. (c) Network depiction of an alternative state in which two species are lost.}
\label{fig_treeshaped}
\end{figure}

Mutualism is defined as an interaction between two species from which both benefit \cite{boucher1982ecology, bronstein2015mutualism}. A marvelous example of a mutualistic interaction can be observed in almost every garden or park, and on many balconies, a bee (or another flying insect) which visits flowers to collect pollen or drink nectar and thereby facilitates the pollination of the visited plants, i.e., the pollinator receives food (+), the plant the service of pollination (+). If one considers not only the mutualistic interaction between two species but between all plant and pollinator species within an ecosystem, one ends up with a complex mutualistic network in which each node represents either a plant or a pollinator species and each link denotes that the pollinator at the one end of the link has been observed to visit the plant species at the other end (Fig. \ref{fig_treeshaped}(a)).

The network of interactions between plants and pollinators yields non-random complex patterns \cite{jordano2003invariant, bascompte2003nested} whose specifics affect the dynamics and the stability of the corresponding ecosystems \cite{memmott2004tolerance, bastolla2009architecture, thebault2010stability, bascompte2013mutualistic, rohr2014structural}. Importantly, the individual contribution to the overall system functioning and robustness varies widely among species -- e.g., the loss of particular keystone species can trigger extinction cascades \cite{memmott2004tolerance, kaiser2010robustness, vieira2015simple, dominguez2015ranking}. However, just like the importance, also the vulnerability or endangerment of species is distributed unevenly -- which has led to non-random losses of species in the past \cite{potts2010global, burkle2013plant}. Differences in the importance and vulnerability of species can be associated with differences in topological properties of the corresponding nodes \cite{saavedra2011strong, dominguez2015ranking}, which are quite pronounced in mutualistic networks. For instance, mutualistic networks typically show a skewed degree distribution with many low degree nodes (specialists) and fewer high degree nodes (generalists) \cite{jordano2003invariant}. Intuitively, the loss of a specialist seems less critical for the integrity of the whole system but might also be more likely as they depend on few or a sole food or pollination source(s) -- in fact, the endangerment of species with few connections has been highlighted multiple times \cite{james2012disentangling, burkle2013plant}. Moreover, the importance of particular generalists is amplified due to the disassortative mixing within plant-pollinator networks, especially due to the tendency of specialists to interact with generalists \cite{bascompte2003nested, vazquez2004asymmetric}.

The tendency for asymmetric interactions paired with the heterogeneous degree distribution results in a global network structure which can be divided into a more or less densely connected core and a periphery of specialists which are mostly linked to generalists in the core \cite{bascompte2003nested, csermely2013structure, lee2016network, miele2020core} (Fig. \ref{fig_treeshaped}(a)). A simple approach to capture this core-periphery division is the $k$-shell decomposition \cite{seidman1983network} which assigns nodes to different shells based on the following procedure: All nodes of degree $k = k_s=1$ are recursively removed and assigned to the $1$-shell until no such nodes remain. To create the higher order shells, $k_s$ is then stepwise increased and the procedure -- recursively remove nodes with degree $k\leq k_s$ and assign them to the $k_s$-shell -- is repeated until each node has been assigned to a shell. Accordingly, nodes in the inner core receive a high shell index, while the most peripheral species are in the $1$-shell (e.g., all specialists and the two nodes in the tree-shaped structure in Fig. \ref{fig_treeshaped}(a)). In a recent theoretical study, Morone et al. \cite{morone2019k} highlighted the significance of the core-periphery distribution of plant-pollinator networks. Using the $k$-shell decomposition, they showcased that the gradual collapse of a mutualistic network follows the core-periphery distribution, with nodes/species in the outermost periphery being lost first while the loss of nodes in the innermost core signals the advent of a complete system collapse.

The work by Morone et al. \cite{morone2019k} is part of a series of recent theoretical works examining the robustness of mutualistic networks in the light of changing environmental conditions \cite{saavedra2013estimating, lever2014sudden, dakos2014critical, jiang2018predicting, lever2020foreseeing, aparicio2021structure} -- e.g., due to increasing anthropogenic stress \cite{potts2010global, goulson2015bee, wagner2021insect}. Studying corresponding scenarios is valuable for understanding, predicting and potentially countering the systemic response of mutualistic networks to the ongoing environmental deterioration \cite{bascompte2023resilience} -- an aspect which is of uttermost importance in view of current pollinator declines \cite{sanchez2019worldwide, cardoso2020scientists}. Importantly, mutualistic networks represent nonlinear systems which, due to the positive interactions between species, inevitably involve positive feedbacks \cite{lever2014sudden, latty2019risk}. Positive feedbacks can reinforce the impact of an initial change and thus allow for an abrupt response to the gradual environmental change. The most severe instance is a so-called tipping point \cite{scheffer2001catastrophic}, which, if approached due to a parameter surpassing a critical threshold, leads to a partial or even complete system collapse. Important insights provided by tipping-related studies are that specialists or generally species in the periphery of mutualistic networks are the ones being affected first as conditions are becoming critical \cite{lever2014sudden, dakos2014critical, morone2019k}, while central species and species in the core are most essential for the integrity of the whole system \cite{morone2019k, jiang2018predicting}.

In a scenario of smoothly and slowly changing conditions, a system can only tip or collapse if a bifurcation is approached at which the present desired state is replaced by an alternative undesired one. However, not only the system response but also disturbances can be large and occur abruptly. Prominent examples of such shock perturbations are extreme events like floods, dry periods or wildfires \cite{coumou2012decade} -- all of which represent extreme stressors for populations \cite{goulson2015bee, soroye2020climate, harvey2020climate, halsch2021insects, wagner2021window}. In mutualistic networks, the occurrence of large shock perturbations is particularly significant as the characteristic positive feedbacks are likely to induce multistability \cite{latty2019risk} -- especially in systems in which the mutualism is obligatory \cite{feng2018unifying}. Multistability denotes the situation that the phase space is populated by multiple coexisting basins of attraction (Fig. \ref{fig_treeshaped}(b)), where each basin comprises all initial conditions which lead to the same attractor or stable long-term behavior. In plant-pollinator networks, one of these attractors can be considered the desired state in which all species coexist ($\mathbf{X_0}$ in Fig. \ref{fig_treeshaped}), while all others are considered undesired states in which some species are lost or extinct ($\mathbf{X_a}$ in Fig. \ref{fig_treeshaped}). Accordingly, a shock perturbation can be fatal if it pushes the system into the basin of any undesired state. As such shock-induced transitions can occur long before a bifurcation point, no prior trend precedes the event, no early warning signals can be detected \cite{scheffer2009early, dakos2014critical} and its consequences are hard to predict \cite{lever2020foreseeing}.

In an earlier study, we determined the most efficient way to trigger such a shock-induced transition from the basin of the desired to the basin of an undesired state -- which we referred to as the \textit{minimal fatal shock} -- for different plant-pollinator networks \cite{halekotte2020minimal}. We found that motifs which turned out to be particularly vulnerable involved links between multiple peripheral nodes and thus are in conflict with the general core-periphery architecture of mutualistic networks. The most outstanding instances of such peripheral motifs are tree-shaped or tree-like structures (Fig. \ref{fig_treeshaped}(a)). A tree-shaped structure \cite{nitzbon2017deciphering} can be defined as a connected subgraph which is located entirely in the $1$-shell and thus, if cut from the rest of the network, fulfills the definition of a tree (it contains no circles), while a tree-like structure is one which resembles a tree-shaped structure. Accordingly, already our former findings suggest that the catchphrase of this work -- \textit{keep the bees off the trees} -- might be appropriate.

However, the consideration of single specific perturbations -- like the minimal fatal shock -- alone does only provide limited insights concerning the endangerment of mutualistic species. One reason is that they only concern a small set of particularly vulnerable species and thus miss to provide any information on less endangered species. Furthermore, it seems unlikely that catastrophic events somehow follow an optimization procedure which would lead to a resemblance of the most efficient disturbance. In this work, we avoid both of these issues by (1) providing information on the endangerment of all species and by (2) considering arbitrary (random) instead of specific perturbations. To this end, we consider plant-pollinator networks as nonlinear dynamical systems (Sec. \ref{basics_model}) exhibiting multistability -- with one desired and multiple undesired states. As in our former work \cite{halekotte2020minimal}, we assume that these systems are subject to large abrupt disturbances which directly affect their state variables. More specifically, we examine how likely a species is to go extinct due to a single large shock perturbation (Sec. \ref{basics_perturb}). However, in contrast to our former work \cite{halekotte2020minimal}, we consider random instead of specific perturbations. The use of a set of random perturbations allows us to derive an endangerment ranking -- ranking all pollinator species from most to least prone to getting extinct. In order to capture the relation between a species' endangerment and its position in the network, we compare the endangerment ranking to rankings which are derived from structural properties of the underlying network topology -- i.e., rankings which are based on centrality metrics (Sec. \ref{basics_central}). In this context, we aim for ranking algorithms which are ecologically meaningful as well as easy to interpret -- in particular, we apply centrality metrics which take into account the degree of specialization, the core-periphery structure of plant-pollinator networks and/or the mutual enforcement between mutualistic partners.

In the remainder of this work, we proceed as follows: We first illustrate the general idea of our work and introduce the necessary tools for our analysis, which include the dynamical model of mutualism, the procedure for creating the endangerment rankings and the selection of applied topological ranking algorithms (Sec. \ref{the_basics}). By applying these tools to different plant-pollinator networks, we then test which centrality metric is best suited to reflect the distribution of species endangerment (Sec. \ref{results_ranking}). Afterwards, we investigate how changing conditions might affect this distribution, first in our standard system setup (Sec. \ref{results_robustness}) and then in an alternative setup which involves a non-trivial topology for the competition between species (Sec. \ref{explicit_comp}). In this context, we also use a composite of different centrality metrics to illustrate how certain topological traits of network nodes contribute to the endangerment of the corresponding species. Finally, we conclude with a discussion of our findings (Sec. \ref{discussion}).

\section{The basics} \label{the_basics}

\begin{figure*}[htb]
\centering
\includegraphics[width=0.7\linewidth]{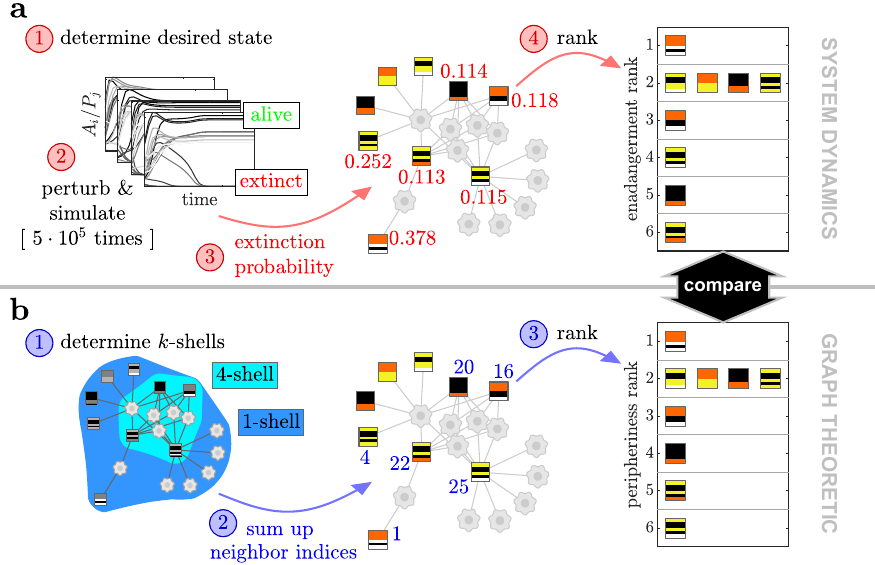}
\caption{\textbf{Creation of endangerment and peripheriness rankings.} (a) Upper half of the figure illustrates the creation of the endangerment ranking (red numbering): (1) Set up dynamical system and determine its desired state in which all species coexist; (2) Perturb desired state and simulate system until final stable state is reached, repeat this $N_{\Delta}$ times; (3) Extinction probability for each species is obtained by counting the number of final states in which the species is extinct and dividing it by the total number of simulations $N_{\Delta}$; (4) Rank species in descending order of extinction probability. (b) Lower half of the figure illustrates the creation of an exemplary peripheriness ranking based on the neighborhood coreness (blue numbering): (1) Determine $k$-shell index for each node; (2)  Neighborhood coreness is obtained as the sum of all neighbor indices; (3) Rank nodes in ascending order of neighborhood coreness. Finally, endangerment ranking and peripheriness ranking are compared.}
\label{fig_centralities}
\end{figure*}

The basic idea of this work is quite simple. In short, we rank the pollinator species within plant-pollinator systems according to their probability of getting extinct (endangerment ranking, Fig. \ref{fig_centralities}(a)) and according to their position within the corresponding network (peripheriness ranking, Fig. \ref{fig_centralities}(b)), and then compare those rankings. In the following, we present the essential cornerstones which we need to obtain the different rankings, (1) for the endangerment ranking: the dynamical model of the plant-pollinator system (Sec. \ref{basics_model}) and the applied perturbation/simulation scheme (Sec. \ref{basics_perturb}), and, (2) for the peripheriness ranking: an overview over the selected ranking algorithms (Sec. \ref{basics_central}).

\subsection{Model of plant-pollinator networks} \label{basics_model}
We consider a simple model of a mutualistic network \cite{bastolla2009architecture, lever2014sudden} in which the dynamics of each species, plant $P$ or animal pollinator $A$, are captured in a differential equation including a term for the intrinsic dynamics $f_i$, the interspecific competition $g_i$ and the mutualistic interaction $m_i$. The latter is combined with an Allee effect $q_i$. For the abundance of an exemplary plant species $P_i$, the dynamics read 
\begin{equation} \label{Eq_mutualism}
    \frac{\mathrm{d} P_i}{\mathrm{d} t} \, = \, \left[ \, f_i(P_i) \,-\, g_i(\mathbf{P}) \,+\, q_i(P_i) \, m_i(\mathbf{A}) \,\right]\, P_i \; ,
\end{equation}while the equation for the abundance of an animal species $A_j$ can be written in the same way by interchanging $P$ and $A$. In the model, the vector $\mathbf{P}$ holds the state variables corresponding to the abundances of all $N_P$ plant species $P_i$ ($i=1, ..., N_P$) and the vector $\mathbf{A}$ holds the abundances of all $N_A$ animal species $A_j$ ($j=1, ..., N_A$).

A species' intrinsic dynamics $f_i$ constitute the dynamics which are independent of any other species, with
\begin{equation}
    f_i(P_i) \, = \, \alpha_i \, - \, \beta^P_{ii} P_i \; .
\end{equation}
In order to obtain multistability \cite{feng2018unifying}, we assume the benefit a species gains from mutualistic interactions (the pollination process) to be obligatory for its own growth and thus we choose the intrinsic net growth rate $\alpha \leq 0$. The strength of the intraspecific competition between individuals of the same species is given by the parameter $\beta^P_{ii}$.

In addition to intraspecific competition, we include the interspecific competition $g_i$ between species from the same class of species (or guild) -- plants compete with plants and animals with animals (intra-guild competition) -- with
\begin{equation} \label{Eq_comp} 
    g_i(\mathbf{P}) \, = \, \sum_{l \neq i}^{N_P} \beta^P_{il} P_l \; .
\end{equation}
In the case of plant species, $\beta^P_{il}$ holds the competitive pressure of plant species $l$ on plant species $i$.

Ultimately, the plant-pollinator network (e.g., Fig. \ref{fig_treeshaped}(a)) is built by the mutualistic interactions. In the most commonly applied formulation -- introduced in \cite{bastolla2009architecture} -- the mutualistic benefit $m_i$ a species obtains from its partners saturates in accordance with a Holling type II functional response
\begin{equation} \label{eq_m}
    m_i(\mathbf{A}) \, = \, \frac{\sum_{j=1}^{N_A} \gamma^P_{ij} A_j}{1 + h \sum_{j=1}^{N_A} \gamma^P_{ij} A_j} \; ,
\end{equation}
where the parameter $h$ describes the handling time and $\gamma$ holds the topology of the mutualistic network and the strength of the involved interactions. The mutualistic network is bipartite and thus the mutualistic benefit of plant species $i$ depends on the abundance of the animals $A_j$ it interacts with (inter-guild mutualism). More specifically, $\gamma^P_{ij}$ gives the relative benefit plant species $i$ obtains from animal species $j$. In accordance with former studies \cite{saavedra2013estimating, lever2014sudden}, we assume that this relative benefit depends on a species' degree of specialization, which is expressed by the following trade-off
\begin{equation} \label{eq_gamma}
\gamma_{ij} \, = \, \gamma_0 \frac{\delta_{ij}}{k_i^{\zeta}} \; ,  
\end{equation}
where $\gamma_0$ is a constant capturing the general mutualistic strength, $\delta_{ij}=1$ if species $i$ and $j$ are linked and $\delta_{ij}=0$ if they are not, $k_i$ is the number of links species $i$ has (its degree) and $\zeta$ specifies the strength of the trade-off ($\zeta \in [0,1]$, with no trade-off for $\zeta=0$ and full trade-off for $\zeta=1$).

Except for the Allee effect $q(P_i)$, Eq. (\ref{Eq_mutualism}) - Eq. (\ref{eq_gamma}) correspond to a commonly used formulation of a mutualistic system (e.g., \cite{bastolla2009architecture, lever2014sudden, halekotte2020minimal}). The Allee effect describes the phenomenon that a small abundance can adversely affect a species' own per-capita growth \cite{courchamp2008allee}. Pollinators and pollinated plants are both likely to be subject to such an effect due to, e.g., inbreeding depression, pollen scarcity, sterilization due to haplodiplocity, or impaired mate finding or cooperation-based defense strategies \cite{darvill2006population, berec2007multiple, busch2008evolution, latty2019risk}. We consider a formulation of the Allee effect which affects the mutualism-dependent growth rate and which is often associated with the issue of mate-finding (or, in the case of plants, attracting pollinators)
\begin{equation} \label{Eq_allee}
    q_i(P_i) \, = \, 1 \, - \, \exp \left(- \frac{P_i}{\theta_i} \right) \; ,
\end{equation}
where the parameter $\theta_i$ controls the strength of the Allee effect. It is important to note that $q_i(P_i) \in [0,1]$ can only weaken but not reverse the mutualistic benefit ($q_i m_i \geq 0$). In combination with the negative intrinsic growth rate $\alpha_i$, this induces an Allee threshold, which means that a species' overall per capita growth rate becomes negative below a certain critical population size ($\mathrm{d} P_i/\mathrm{d} t < 0$ for $P_i < P^{crit}_i$).

\subsection{Extinction probability and endangerment rank} \label{basics_perturb}
The premise of this study is that the dynamical system of plants and pollinators exhibits multistability and that one desired state $\mathbf{X_0}$ (Fig. \ref{fig_treeshaped}(ab)) in which all species coexist competes with multiple undesired states in which some species are lost or extinct (e.g., $\mathbf{X_a}$ in Fig. \ref{fig_treeshaped}(bc)). Accordingly, a perturbation can induce the loss of species if it pushes the system state into the basin of attraction of an undesired state (Fig. \ref{fig_treeshaped}(b)). This assumption is essential since we assume that the desired state is locally attractive at all times but that species are endangered due to singular large perturbations which can be interpreted as unspecific extreme events. 

Our aim is to assign an endangerment score and rank to each species in the light of such large shock perturbations. In order to do this, we need to make some assumptions regarding the perturbations. First of all, we consider single perturbations which directly affect the state variables and thus the position of the system in phase space (Fig. \ref{fig_treeshaped}(b)). Moreover, we assume that the perturbations are random and solely reduce the abundances of species (parameters are not affected). For the sake of simplicity, we apply the most simple perturbation scheme by drawing the initial abundance of each species from a uniform distribution within the interval $[0, N^*_i]$ (all species are affected), where $N^*_i$ is the abundance of species $i$ in the desired steady state $\mathbf{X_0}$. The ultimate effect of a perturbation is evaluated by evolving the system until it reaches its final stationary state (by numerical integration \cite{ansmann2018efficiently}), which can be the desired or an undesired state.

In the end, we determine the endangerment ranking in four steps (see Fig. \ref{fig_centralities}(a)): (1) We numerically determine the desired state of the system. (2) We perturb the desired state a number of times $N_{\Delta}$ and determine the corresponding final stationary states. (3) For each species $i$, we count in how many $L_i$ of the approached states it is extinct, and obtain the extinction probability as $\Omega_i = L_i/N_{\Delta}$. (4) We then rank the species according to their extinction probabilities $\Omega$ from most to least endangered, where the species with the highest $\Omega_i$ obtains the rank $1$.

\subsection{Centrality metrics and ranking algorithms} \label{basics_central}
Topological ranking algorithms or centrality metrics usually aim at sorting nodes of a network with regard to their 'importance', often with emphasis on identifying particularly important nodes \cite{lu2016vital, liao2017ranking}. In this work, we test different ranking algorithms with regard to their ability to rank nodes of mutualistic networks in accordance with the endangerment of the corresponding species (see Fig. \ref{fig_centralities}). Since we assume that the most endangered species are located in the network periphery, we always rank the species from most to least peripheral (or from least to most central) and refer to the obtained sorting as peripheriness ranking (see Fig. \ref{fig_centralities}(b)). In general, we aim for metrics which are in some way ecologically meaningful as well as easy to interpret.

\textbf{degree:} The simplest measure of centrality is the \textit{degree} ($k$) which is defined as the number of links a node has. As it is easy to access and interpret, the degree is often a good choice. For instance, in a mutualistic network of plants and pollinators, the degree is equivalent to the degree of specialization of species. Since the particular endangerment of specialist species has been emphasized by multiple empirical studies, the degree might be a suitable starting point for linking network topology and endangerment.

\textbf{Iterative refinement:} However, the degree is a purely local metric which neglects the importance or centrality of neighbors. In a mutualistic system, each species depends on partner species which in turn rely on further other partners, which makes networks of mutualistic interdependencies prime examples of mutual reinforcement. Iterative refinement metrics like the \textit{eigenvector centrality} (EV) \cite{bonacich1987power} consider this aspect as they assign each node a score which depends on the number and scores of its neighbors. Due to its self-referential definition, which reads\begin{equation}
    \mathbf{x} = \kappa^{-1} \mathbf{B} \mathbf{x} \; ,
\end{equation}
the eigenvector centrality $\mathbf{x}$ takes into account the whole network structure -- contained in the adjacency matrix $\mathbf{B}$. In this notation, the constant $\kappa$ corresponds to the largest eigenvalue of $\mathbf{B}$.

One disadvantage of the eigenvector centrality is that it tends to accumulate in a few nodes. In order to avoid this, we introduce a saturation to the recursive definition of the eigenvector centrality
\begin{equation}
    \mathbf{x} = \frac{ \mathbf{B} \mathbf{x}}{1 + h_{e} \mathbf{B} \mathbf{x}} \;.
\end{equation}
The chosen saturation is in line with the typical formulation of mutualistic benefits which is often described by a Holling type II functional response (see Eq. (\ref{eq_m})). In reference to the Holling type, we name this centrality index the \textit{eigenvector centrality type II} (EV2). In the framework of Holling, the parameter $h_e$ corresponds to the handling time. For the sake of simplicity, we set $h_e=1$ throughout this work.

\textbf{$\mathbf{k}$-shell index:} An alternative and simple approach for considering a node's position within the potentially hierarchical structure of a network is the $k$-shell decomposition \cite{seidman1983network}, which assigns a $k$\textit{-shell index} ($k_s$) to each node (see Sec. \ref{introduction}). In fact, since mutualistic networks possess a core-periphery structure which has been shown to be essential for their structural integrity \cite{morone2019k}, a ranking considering the coreness (or peripheriness) of a node might be worthwhile.

\textbf{$\mathbf{k}$-shell refinement:} Because many nodes are assigned to the same shell, the $k$-shell decomposition is fairly limited when used to rank all nodes of a network (see left side of Fig. \ref{fig_centralities}(b)). However, in plant-pollinator networks, which typically include many specialists, the same is true for the degree. Fortunately, owing to its generally good performance, especially in capturing influential spreaders \cite{kitsak2010identification}, multiple adaptations of the $k$-shell decomposition have been proposed (e.g., \cite{zeng2013ranking, bae2014identifying, liu2016locating, ma2016identifying}). All of these adaptations provide an improved resolution concerning the distinction of nodes. 

In this work, we apply two of these adaptations. The first is the \textit{neighborhood coreness} ($C_{nc}$) \cite{bae2014identifying} which is obtained by simply summing up the $k$-shell indices of a node's direct neighbors $\Lambda_1$
\begin{equation}
C_{nc} = \sum_{j \in \Lambda_1} k_s(j) \; .
\end{equation} Importantly, this yields an index which combines the degree of a node with the coreness of its neighbors -- a rather basic approach to combine connectedness with position. In Fig. \ref{fig_centralities}(b), this index is used to demonstrate the general procedure to obtain a peripheriness ranking. 

The second adaptation -- the \textit{neighborhood centrality} ($C_N$) \cite{liu2016identify} -- is slightly more elaborated but still intuitively interpretable (see below). It is defined as
\begin{equation}
    C_N = \sum_{r=0}^{R} \left( d^r \sum_{j \in \Lambda_r} \theta_j \right) \; ,
\end{equation} where $\theta$ is the benchmark centrality (in our case, the $k$-shell index) and $\Lambda_r$ is the set of nodes whose distance to node $i$ equals $r$. The parameter $R$ denotes the range of neighborhood which is taken into account, i.e., for $R=1$, only a node's direct neighbors are considered, for $R=2$, also the next order neighbors and so on. The parameter $d$ controls the diminishing impact of nodes in the neighborhood of node $i$ with increasing distance to node $i$. According to Liu et al. \cite{liu2016identify}, $d$ should be in the range $[0, 1]$. 

Using the $k$-shell index as the benchmark centrality and setting the parameter $d$ to an arbitrary but very small value ($d=0.001$) allows an instructive interpretation of the neighborhood centrality. For $d\ll1$, the rank of a node is primarily based on its own $k$-shell ($r=0$), while the sum of the $k$-shell indices of its neighbors is a secondary argument ($r=1$), the sum of the $k$-shell indices of its second-order neighbors a tertiary argument ($r=2$) and so on (we set $R=2$ and thus only consider primary, secondary and tertiary arguments). Due to this hierarchical configuration of arguments (most, second most, third most important), the neighborhood centrality highlights nodes within tree-shaped structures. In fact, since both their own and the $k$-shell indices of their neighbors are $1$, nodes in the outermost positions of tree-shaped structures obtain the lowest centralities.

\section{Species in Trees are most endangered} \label{results_ranking}
Now that we have established the necessary tools for creating endangerment and peripheriness rankings, we are set to examine how the two relate. Accordingly, what follows is the centerpiece of this work in which we present the results which provide the grounding for the theme of this work (\textit{keep the bees off the trees}).

\subsection{Exemplary endangerment ranking}
\begin{figure}[ht]
\centering
\includegraphics[width=1.0\linewidth]{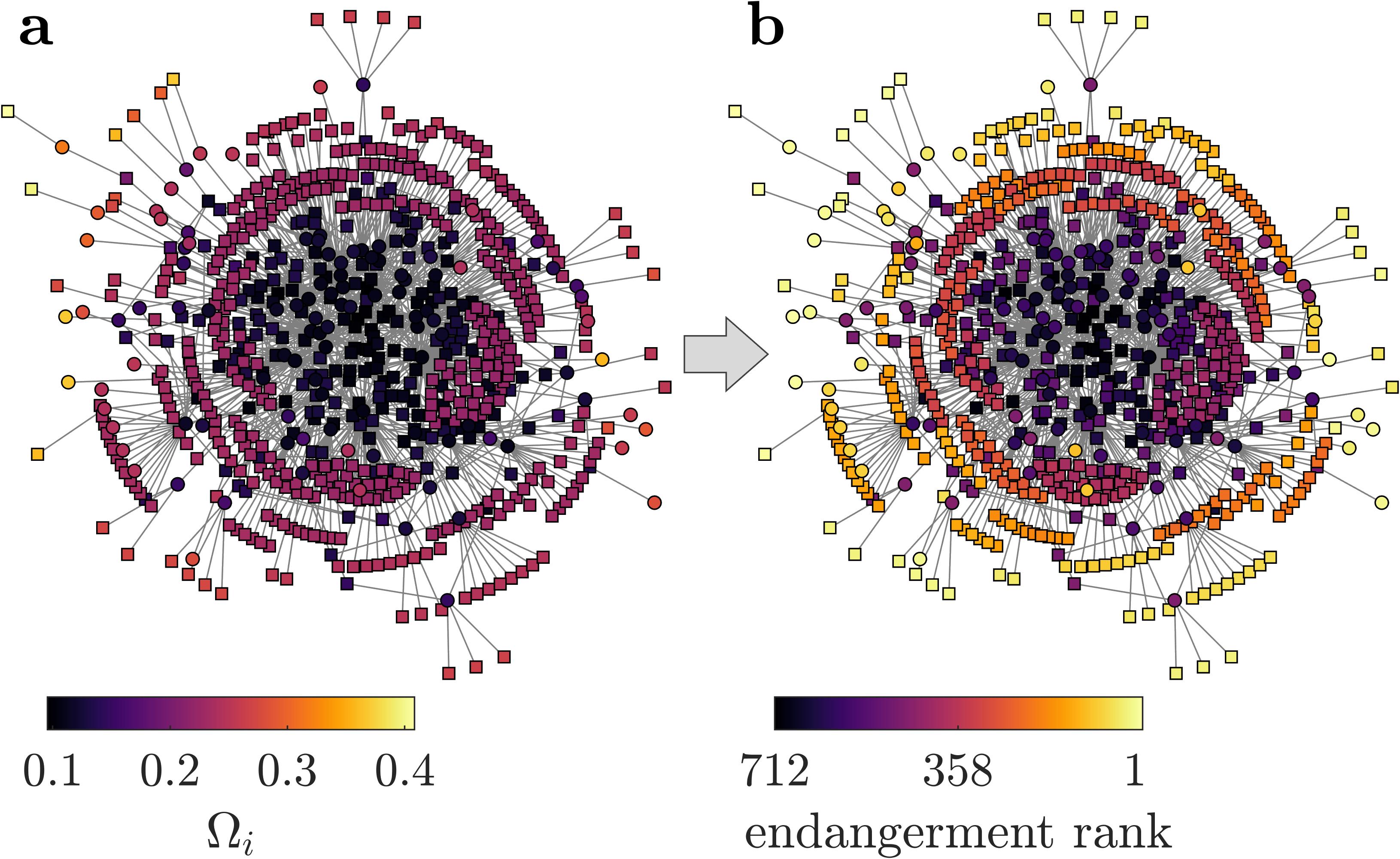}
\caption{\textbf{Endangerment of species within an exemplary network.} (a) Extinction probabilities $\Omega$ of species. (b) Endangerment ranks of species. Squares depict animals and circles plants. Parametrization corresponds to the standard setup with competition type I (see Appendix \ref{appendix_parameters}). Placement of the vertices is based on the Kamada-Kawai algorithm \cite{kamada1989algorithm}.}
\label{fig_2nets}
\end{figure}

First, we consider an exemplary mutualistic system (Fig. \ref{fig_2nets}) which stems from a plant-pollinator network on the Amami Islands in the Ryukyu Archipelago, Japan \cite{kato2000anthophilous} (M\_PL\_044 in Table \ref{table1}). To start with, we set up the dynamical system according to the standard parametrization (see Sec. \ref{basics_model} and Appendix \ref{appendix_parameters}) and numerically determine the desired state in which all species coexist ($P^*_i>0 \;\forall\; i$ and $A^*_j>0 \;\forall\; j$). Based on this pre-disturbance state, we then calculate the extinction probabilities $\Omega$ for all species and derive the corresponding endangerment ranking (according to the explanations in Fig. \ref{fig_centralities}(a) and in Sec. \ref{basics_perturb} with $N_{\Delta}=5 \times 10^5$).

Already the visual inspection of $\Omega$ (Fig. \ref{fig_2nets}(a)) allows the conclusion that the degree of specialization and its position within the core-periphery structure determine a species' endangerment. Especially a low degree which corresponds to a high level of specialization seems to establish a species' status as being endangered. However, the endangerment ranking (Fig. \ref{fig_2nets}(b)) reveals that the core-periphery structure shapes the endangerment ranking of species as well. We find that specialists close to the core hold a lower endangerment rank than species in the outer parts of the network (according to the Kamada-Kawai visualization). Moreover, we find that the most endangered species are in fact the ones in the outermost positions of tree-like structures (\textit{bees in trees}).

\subsection{Peripheriness vs. endangerment rankings} \label{sec_cen_vs_end}
In the following, we compare the endangerment rankings to rankings based on different centrality metrics (see Sec. \ref{basics_central} and Fig. \ref{fig_centralities}(b)) for a set consisting of 11 of the largest plant-pollinator networks, including the exemplary network, from the web of life database (see Appendix \ref{appendix_weboflife}). However, we sort species from least to most central (reversed centrality or peripheriness rankings) as we assume the correlation between centrality and endangerment to be negative (e.g., low degree equals high endangerment). Importantly, we do not consider all nodes of a network. The reason for this is twofold. First of all, species with the same set of neighbors are topologically and dynamically equivalent (due to our parametrization, see Appendix \ref{appendix_parameters}). We therefore only include one node from each set of equivalent nodes in the comparison. Furthermore, comparing the endangerment of animals and plants is questionable since their competitive terms are separated. We concentrate on the numerically dominant group, the pollinators -- the number of unique animal species is denoted as $\tilde{N}_A$.

To evaluate the goodness of fit of the rankings obtained by different centrality indices and the endangerment in these networks (i.e., to "compare" endangerment and peripheriness rankings, see Fig. \ref{fig_centralities}), we propose two measures. The first is Kendall's tau \cite{kendall1938new} which represents a common tool (e.g., \cite{bae2014identifying, liu2016identify, ma2016identifying, liu2016locating, ahajjam2018identification}) to quantify the overall rank correlation between two metrics -- one being the reversed centrality index $x_i$ and the other the endangerment $y_i$ in our case. The rank correlation coefficient $\tau$ is defined as \begin{equation}
    \tau = \frac{2}{n(n-1)} \sum_{i<j} \text{sign}[(x_i-x_j)(y_i-y_j)] \; ,
\end{equation}
where $n$ is the number of nodes and $\text{sign[z]}$ is the sign-function which gives $\text{sign[z]} = +1$ if $z>0$ and $\text{sign[z]} = -1$ if $z<0$. A value of $\tau$ close to 1 indicates a strong positive correlation between the two rankings, which means a high accuracy of the proposed index. 

\begin{figure}[ht]
\centering
\includegraphics[width=1.0\linewidth]{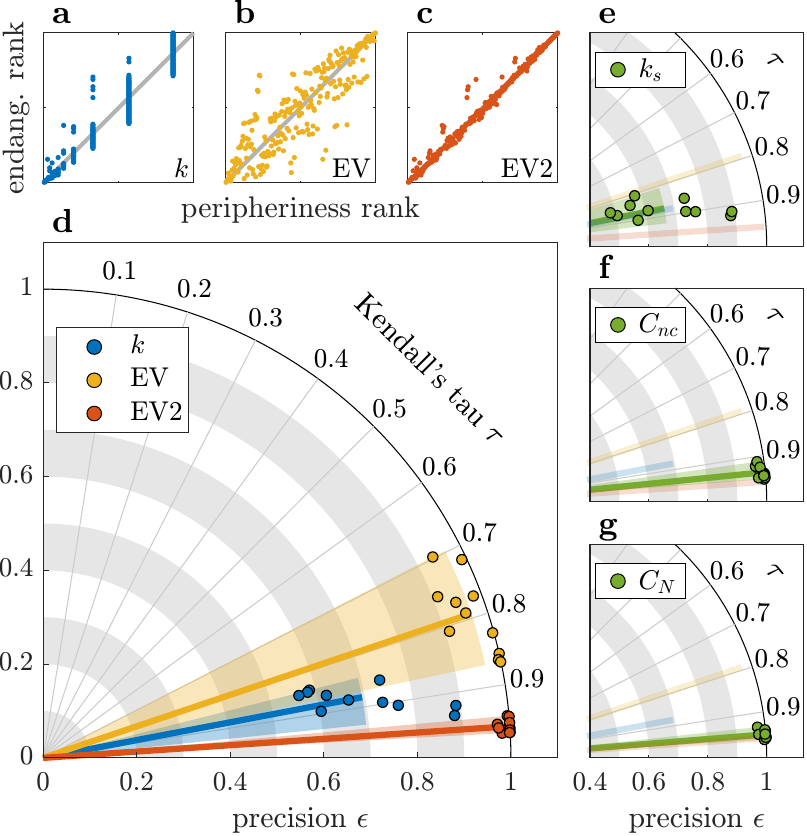}
\caption{\textbf{Rank correlation between reversed centrality indices and the endangerment of species.} (a-c) Endangerment rank versus peripheriness (inverse centrality) rank based on degree $k$ (a), eigenvector centrality EV (b) and eigenvector centrality type II EV2 (c) for an exemplary network. (d) Rank correlation coefficient based on Kendall's tau $\tau$ (angle) and precision $\epsilon$ (radius) for 11 empirical plant-pollinator networks for degree (blue), eigenvector centrality (yellow) and eigenvector centrality type II (orange). Every dot corresponds to the correlation coefficient and precision for one network, the line depicts the mean value of all networks and the fan the range of correlation coefficients. (e-g) Kendall's tau $\tau$ and precision $\epsilon$ for $k$-shell index $k_s$ (e), neighborhood coreness $C_{nc}$ (f) and neighborhood centrality $C_N$ (g) -- each depicted in green. For comparison, we added the mean Kendall's tau and precision for degree (blue), EV (yellow) and EV2 (orange).}
\label{fig_allnetworks}
\end{figure}

The second measure is inspired by the imprecision function proposed by Kitsak et al. \cite{kitsak2010identification}. In our adaptation, we compare the average extinction probability $M_{cen}(\rho)$ of a small fraction $\rho \tilde{N}_A$ ($0<\rho<1$) of all considered nodes/pollinators $\tilde{N}_A$ to the average extinction probability $M_{most}(\rho)$ of the $q\tilde{N}_A$ nodes with the highest extinction probability $\Omega_i$. It should be noted that if the set is not unique (e.g., due to multiple nodes having the same degree), $M_{cen}(\rho)$ is calculated as the worst possible realization. The measure is then obtained as
\begin{equation}
\epsilon \; =\;  \frac{M_{cen}(\rho) - \Omega_{min}}{M_{most}(\rho) - \Omega_{min}} \; , 
\end{equation}
where the minimal extinction probability $\Omega_{min}$ of all $\tilde{N}_A$ nodes is used as a scaling factor. The closer $\epsilon$ is to $1$, the better the corresponding reversed centrality index is in correctly capturing the most severely endangered species. Accordingly, we refer to $\epsilon$ as the precision function (imprecision would be $1-\epsilon$, see \cite{kitsak2010identification}). The precision function puts special emphasis on a small subset of the ranking (we choose $\rho=0.05$) which represents the few most endangered species. In this regard, the precision complements Kendall's tau. In Fig. \ref{fig_allnetworks}, both measures -- $\tau$ (angle) and $\epsilon$ (radius) -- are displayed together.

The degree (blue), the eigenvector centrality (yellow) and the eigenvector centrality type II (orange) all provide at least a decent approximation of the endangerment ranking (Fig. \ref{fig_allnetworks}(a-d)). However, the results also reveal some significant differences between the three ranking algorithms. First of all, the impression that the degree is a good indicator for the endangerment of a species is backed up by the general strong correlation between the rankings (high $\tau$ for blue marks in Fig. \ref{fig_allnetworks}(d)). Noteworthy is that the strong impact of the degree is in part due to the Allee effect which in combination with large perturbations induces a certain ground-endangerment for every species. As any species can be lost if its own abundance falls below a certain threshold, species with low degree have a much higher chance of losing all partners and thus eventually becoming extinct (see Appendix \ref{appendix_allee} for an elaborated explanation). However, the degree is not capable of separating the set of the most endangered species (low $\epsilon$ for blue marks in Fig. \ref{fig_allnetworks}(d)). In fact, the degree 'only' provides a kind of presorting which puts species into certain classes of endangerment, with specialists being in the first class of endangerment (see Fig. \ref{fig_allnetworks}(a)).

By contrast, the eigenvector centrality EV (Fig. \ref{fig_allnetworks}(b) and yellow marks in \ref{fig_allnetworks}(d)) captures the most endangered species well (high $\epsilon$) but provides a worse overall ranking than the degree (lower $\tau$). Finally, the best agreement with the endangerment ranking is provided by the eigenvector centrality type II EV2 (Fig. \ref{fig_allnetworks}(c) and orange marks in Fig. \ref{fig_allnetworks}(d)) which surpasses the other two in both the overall rank correlation ($\tau$) and the capability of identifying the most severely endangered species ($\epsilon$). This is because the own connectiveness (degree) and the position of its neighbor(s), which both significantly contribute to the EV2, determine the endangerment of a species. The contribution of both aspects can best be demonstrated by means of the mutualistic benefit a species obtains right after the initial perturbation which depends on the number and the pre-disturbance abundance of partner species. As the latter arises from the mutual enforcement between species, it strongly depends on the partners' position within the network (see Appendix \ref{appendix_irwinhall} for an elaborated explanation).

In order to obtain a better insight into the correlation between the endangerment of species and the specific core-periphery structure of mutualistic networks, we consider simple measures of the coreness, the $k$-shell index and two of its variants (see Sec. \ref{basics_central}). The $k$-shell index provides a fit similar to the one achieved by the degree, concerning both the rank correlation $\tau$ and the precision $\epsilon$ (green marks in Fig. \ref{fig_allnetworks}(e)). Accordingly, a node's coreness alone does not provide a sufficient explanation for the particular endangerment of some specialists (this comes as no surprise as a degree of $1$ always implies that the corresponding node is located in the outermost $1$-shell). However, already the simplest adaptation of the $k$-shell index -- the neighborhood coreness $C_{nc}$ (green marks in Fig. \ref{fig_allnetworks}(f)) which is obtained as the sum of a node's neighbors $k$-shell indices -- works pretty well (only slightly worse than the EV2). The reason is that this index combines the two aspects which can be shown to be decisive for a species' vulnerability to random shocks, its degree and the coreness of its neighborhood (see Appendix \ref{appendix_allee} and \ref{appendix_irwinhall}). An even better agreement -- roughly as good as the EV2 -- is obtained by the neighborhood centrality $C_N$ based on the $k$-shell index (green marks Fig. \ref{fig_allnetworks}(g)). As mentioned earlier (see Sec. \ref{basics_central}), this metric is especially beneficial for the interpretation of our results as it uses a node's own, its first-order neighbors' and second-order neighbors' shell indices as primary, secondary and tertiary arguments and thus highlights nodes which are located deep within tree-shaped and tree-like structures as the least central ones. Accordingly, the good fit between the reversed neighborhood centrality and the endangerment encourages our initial claim (and title of this work): \textit{Keep the bees off the trees}.

\subsection{Treeness vs. extinction probability}

\begin{figure}[ht]
\centering
\includegraphics[width=1.0\linewidth]
{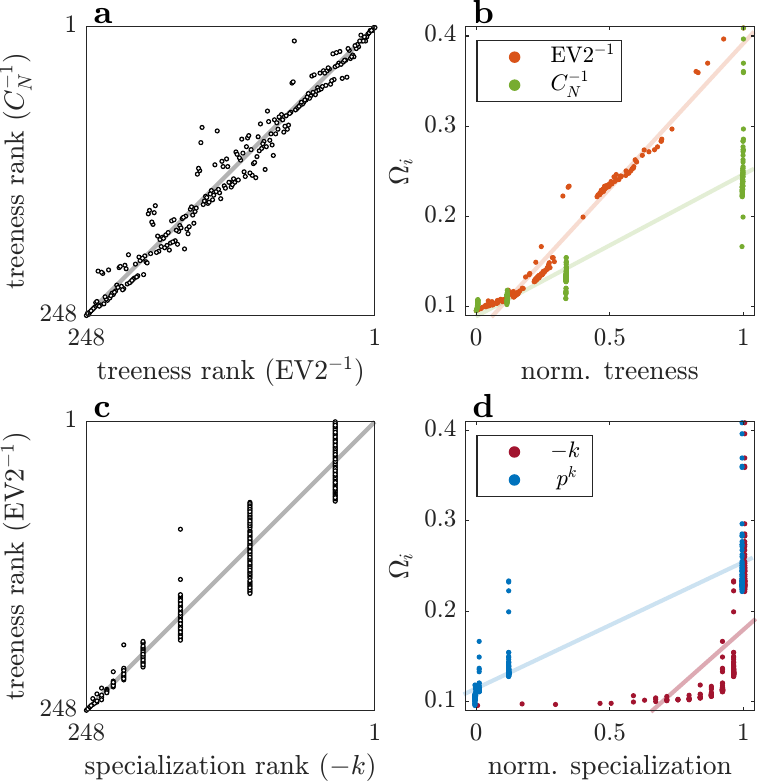}
\caption{\textbf{Relation between eigenvector centrality type II (EV2), neighborhood centrality ($C_N$) and degree ($k$) within an exemplary network.} (a) Correlation between the treeness rankings based on EV2 and based on $C_N$. (b) Linear correlation between the treeness (normalized to the interval $[0, 1]$) based on EV2 and based on $C_N$ and the extinction probabilities $\Omega$. (c) Correlation between the specialization ranking based on the negative degree $-k$ and the treeness ranking based on EV2. (d) Linear correlation between the specialization (normalized to the interval $[0, 1]$) based on $-k$ and based on the power law description $p^k$ and the extinction probabilities $\Omega$. The parameter $p$ is set as the difference between the lowest $\Omega_i$ of all nodes with degree $1$ and the overall lowest $\Omega_i$ (see Appendix \ref{appendix_allee}).}
\label{fig_treeness}
\end{figure}

Both, the eigenvector centrality type II (EV2) and the neighborhood centrality ($C_N$) capture the endangerment ranking of pollinators exceptionally well. This comes as no surprise since the rankings obtained by the two algorithms resemble one another (Fig. \ref{fig_treeness}(a)), which is why, from now on, we refer to both of them as treeness rankings. However, importantly, the underlying distribution of centrality (or better treeness) indices shows immense differences. This becomes apparent when looking at the linear correlation between the two indices and the actual extinction probabilities $\Omega$ (Fig. \ref{fig_treeness}(b)). While the block-wise distribution of $C_N$ indices provides only a very broad approximation to $\Omega$, the EV2 -- which is based on the Holling type II functional response also used in the descriptor of the mutualistic benefit (see Eq. (\ref{eq_m})) -- captures the distribution of extinction probabilities very well.

An important factor to the great fit between the endangerment described by $\Omega$ and the treeness described by the EV2 is that the degree of specialization significantly contributes to the latter metric. In fact, we find that all pollinators which obtain the 69 highest treeness ranks (according to the EV2) have a degree of $k=1$ (see Fig. \ref{fig_treeness}(c)). But, in contrast to the EV2 (see Fig. \ref{fig_treeness}(b)), the degree $k$ shows a very poor linear correlation with the extinction probability $\Omega$ (Fig. \ref{fig_treeness}(d)). The reason is that the impact of the degree on species endangerment is not linear but broadly follows a power law which describes a species' chance of losing all mutualistic partners due to the initial shock perturbation, i.e., $\Omega_i \propto p^{k_i}$ (where the constant $p$ provides some kind of basal-endangerment; see Appendix \ref{appendix_allee} for the derivation of this relation based on a simplified model of the mutualistic system). Therefore, in the following, we will use the power law description of the degree $p^k$ as a measure of a species' specialization. As a measure of the treeness, we will use the eigenvector centrality type II (EV2), since it captures both the endangerment ranking and the distribution of extinction probabilities $\Omega$ particularly well.

\section{Species in trees are most endangered under different conditions} \label{results_robustness}

So far, our claim \textit{keep the bees off the trees} is based on a single parametrization scheme. However, the chosen parameters are somehow arbitrary and, as often claimed in tipping-related studies, environmental conditions change. Accordingly, it is informative to test for the robustness of the obtained rank correlations. To this end, we consider once again the exemplary plant-pollinator network from the Amami Islands (see Fig. \ref{fig_2nets}) and check how the endangerment of species evolves when an exemplary parameter changes.

\subsection{Robustness of the endangerment ranking} \label{vary_alpha}
First, we check the robustness of the endangerment ranking for the standard system setup (see Appendix \ref{appendix_parameters}). Our parameter of choice is the intrinsic growth rate $\alpha$ which we assume to be negative and whose further decrease can be interpreted as a globally increasing stress level affecting all species in the same way (e.g., harsher conditions due to anthropogenic impact). We vary $\alpha$ within an interval in which the mutualism remains obligatory ($\alpha \leq 0$) and in which the long-term coexistence of all species is possible (the desired state is stable). It should, however, be noted that for the most negative values of $\alpha$, the system is close to a bifurcation (tipping point) at which some species would inevitable go extinct.

\begin{figure}[ht]
\centering
\includegraphics[width=1.0\linewidth]{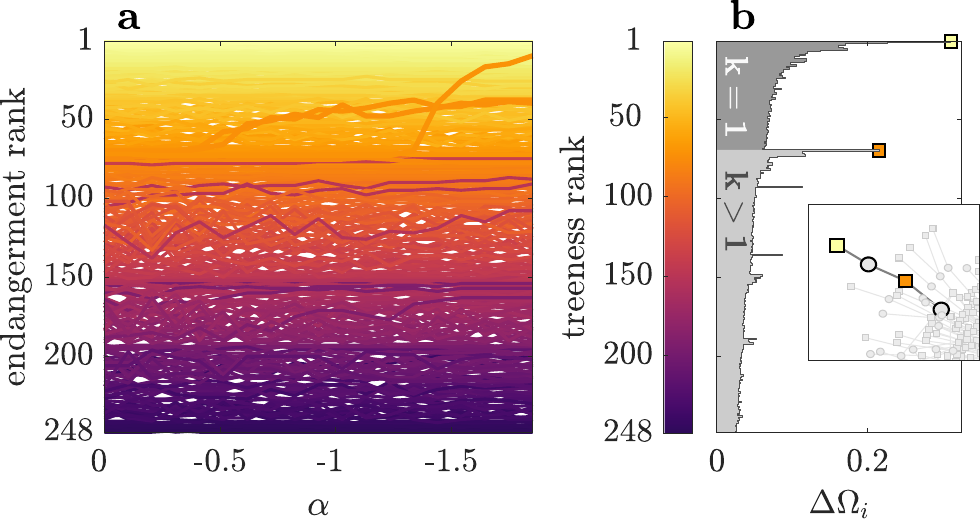}
\caption{\textbf{Evolution of the endangerment ranking with increasing stress level.} (a) Development of endangerment ranks as the intrinsic growth rate $\alpha$ is decreased from $\alpha=0$ to $\alpha \approx -1.7$. Color coding of the different species corresponds to treeness rank based on the EV2 of the respective node (see colorbar in the center). (b) Change in extinction probability $\Delta \Omega$ from highest to lowest $\alpha$, $\Delta \Omega_i = \Omega_i(\alpha \approx -1.7) - \Omega_i(\alpha = 0)$. Species are ordered according to their treeness rank (in line with the axis of the colorbar). Location of two  pollinators/nodes (yellow and orange square) with exceptionally high increase in $\Omega_i$ is highlighted in the small inset.}
\label{fig_twocomps}
\end{figure}

We find that, overall, the endangerment ranking is pretty robust and thus the proposed centrality metrics, like the EV2, provide a good fit for all values of $\alpha$. This can be seen in the consistent distribution of colors in Fig. \ref{fig_twocomps}(a): Nodes in the core (dark/purple) permanently remain at low ranks, while peripheral nodes (bright/yellow) occupy the upper endangerment ranks. Nevertheless, we do observe that for a few individual nodes the endangerment rank changes significantly -- particularly noticeable is "the jump" of one of the orange-colored nodes from endangerment rank $73$ (at $\alpha \approx -1.2$) to rank $10$ (at $\alpha \approx -1.7$, Fig. \ref{fig_twocomps}(a)).

\subsection{"The jump"} 

In order to better understand "the jump" and to check whether it might undermine our claim \textit{keep the bees off the trees}, we take a look at the change in the absolute extinction probabilities $\Delta \Omega$ from the highest to the lowest $\alpha$ (Fig. \ref{fig_twocomps}(b)), $\Delta \Omega_i = \Omega_i(\alpha \approx -1.7) - \Omega_i(\alpha = 0)$. The absolute extinction probabilities $\Omega$ do of course adapt to the changing conditions -- lower growth rates $\alpha$ generally entail higher extinction probabilities $\Omega$ and thus $\Delta \Omega_i>0$ for all species $i$. However, the change $\Delta \Omega$ is not evenly distributed across species (Fig. \ref{fig_twocomps}(b)). We find that more peripheral species (higher treeness ranks) tend to show a higher increase in $\Omega_i$ than species which are closer to the core (lower treeness ranks). But, there are some clear outliers to this trend. The most exposed of them is the above-mentioned node which makes "the jump" (marked by an orange square in Fig. \ref{fig_twocomps}(b)). Taking a look a the corresponding node's position in the network (inset in Fig. \ref{fig_twocomps}(b)), we see that its high $\Delta \Omega_i$ and "the jump" in the endangerment ranking can be associated with the node's proximity to the most peripheral (highest treeness) and most endangered (highest $\Omega_i$) pollinator-node in the system. Furthermore, the node is itself located within a tree-structure. Accordingly, "the jump" does not undermine our general claim \textit{keep the bees off the trees} since the highest endangerment still originates from the most peripheral species.

\subsection{Composite Centrality Index}

What we do observe is that, for $\alpha$ close to $0$, the set of the most endangered species only consists of specialists with a degree of $k=1$ (i.e., there are $69$ unique species with a degree of $1$, all of them are within the set of the $69$ most endangered species). Due to the strong increase in the extinction probability of some species -- as, for instance, the one which makes "the jump" (Fig. \ref{fig_twocomps}) -- this changes and some nodes with degree $k>1$ enter the set of the most endangered species as $\alpha$ decreases (i.e., for smaller $\alpha$, the set of the $69$ most endangered species also contains species with $k>1$).

\begin{figure}[ht]
\centering
\includegraphics[width=0.9\linewidth]{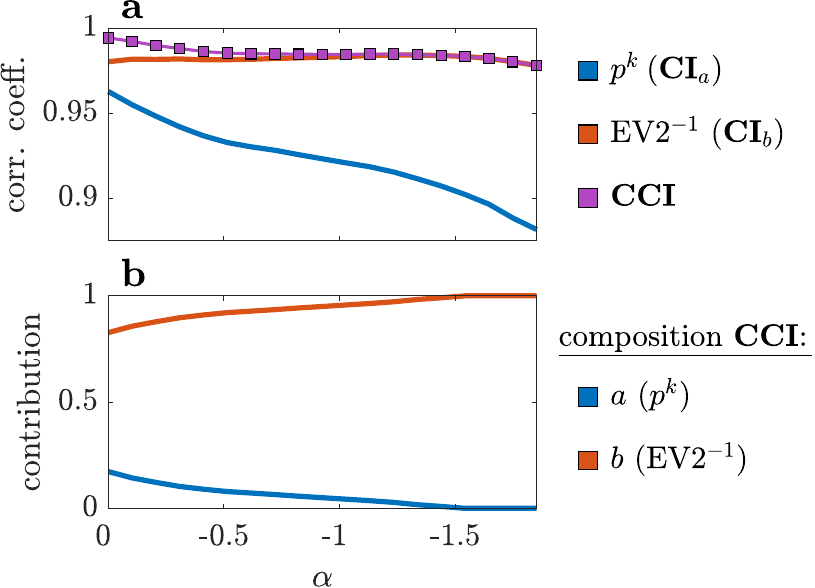}
\caption{\textbf{Topological determinants of endangerment for different values of $\alpha$.} (a) Linear correlation between the extinction probability $\Omega$ and $p^k$ (blue line), EV2 (orange line) and the best composite centrality index $\mathbf{CCI}$ (purple squares) which is obtained as the weighted sum of the normalized forms of $p^k$ and EV2$^{-1}$. (b) Best composition of the $\mathbf{CCI}$ with $a$ describing the contribution of $p^k$ and $b$ the contribution of EV2$^{-1}$.}
\label{fig_strange}
\end{figure}

In fact, by examining the correlation between the extinction probability $\Omega$ and the power law $p^{k}$, we can see that the degree loses explanatory power for determining a species' endangerment. For good conditions (greater $\alpha$), the endangerment of species can well be captured by the power law $p^{k}$, i.e., $\Omega_i \propto p^{k_i}$. However, as $\alpha$ is decreased, the correlation coefficient declines (blue line in Fig. \ref{fig_strange}(a)). By contrast, the EV2 -- which also depends on the degree -- captures the extinction probability $\Omega$ very well for all values of $\alpha$ and only slightly declines for smaller $\alpha$ (red line in Fig. \ref{fig_strange}(a)). 

Although the EV2 provides a good fit for all values of $\alpha$, we test whether the degree can add additional explanatory value. To this end, we apply a hybrid method \cite{erkol2019systematic} by setting up a composite centrality index 
\begin{equation} \label{eq_CCI_one}
    \mathbf{CCI} \; = \; a \, \mathbf{CI}_a \, + \, b \, \mathbf{CI}_b \;\; ,
\end{equation}
with $a+b=1$ and $a,b \ge 0 $. The $\mathbf{CCI}$ is obtained as the weighted sum of two normalized centrality indices, where $\mathbf{CI}_a$ is the normalized form of $p^{k}$ and $\mathbf{CI}_b$ the normalized form of EV2$^{-1}$. The normalization is obtained by setting the length of each vector to $1$. To receive the best composition of $\mathbf{CCI}$, we then check for each value of $\alpha$ which combination of $a$ and $b$ provides the highest Pearson's linear correlation coefficient with the extinction probability $\Omega$.

Following the best composition of $\mathbf{CCI}$ (Fig. \ref{fig_strange}(b)), we see that the EV2 is the dominant descriptor of the endangerment for all $\alpha$. For greater $\alpha$, the EV2 underestimates the explanatory value of the degree and thus a composite which takes a small contribution of the degree provides the highest correlation coefficient with the endangerment. As $\alpha$ decreases the contribution of the degree vanishes, i.e., the importance of the degree for the overall endangerment decreases.

\section{Species in trees are most endangered -- but which trees?} \label{explicit_comp} 

Now, after having established that our claim \textit{keep the bees off the trees} is rather robust against changes in an exemplary parameter, we shortly illuminate an aspect which might undermine its general validity: Real plant-pollinator systems are actually multilayer networks. For instance, in addition to mutualistic links with certain plants, pollinators also have antagonistic links to other pollinators with whom they compete for food and/or nesting sites. 

Due to the consideration of intra-guild competition (Eq. (\ref{Eq_comp})), the mutualistic model in principle already considers the multilayer nature of plant-pollinator networks. However, the assumption that all species within one guild compete with all other species in the same guild in a uniform manner (see left side of Fig. \ref{fig_twocomps_explicit}(a)) -- also called \emph{competition of mean-field type} \cite{bastolla2009architecture} -- represents a strong simplification \cite{gracia2018joint}. Although a realistic topology is difficult to obtain, a reasonable alternative to the mean-field approach is to assume that the resources for which species compete are their mutualistic partners \cite{gracia2018joint}: Plants compete for pollinators and pollinators for plants. Accordingly, we can derive an explicit competition topology by drawing a competitive link between two species from the same guild if they share a common mutualistic partner. Importantly, this way of constructing the competition topology gives rise to a potential benefit of being located in the periphery of the mutualistic network, since peripheral species tend to obtain fewer competitive links than species which are located in the core of the mutualistic network (i.e., being peripheral allows species to avoid competition). By applying this scheme, we obtain a multilayer network which includes explicit topologies for both inter-guild mutualism and intra-guild competition (see right side of Fig. \ref{fig_twocomps_explicit}(a)). In the following, we refer to the corresponding system setup as the \textit{multilayer setup} (see Appendix \ref{appendix_parameters} for the parametrization of this setup) -- in contrast to the \textit{standard setup} which we considered so far.

\subsection{Robustness of the endangerment ranking}

\begin{figure}[ht]
\centering
\includegraphics[width=0.85\linewidth]{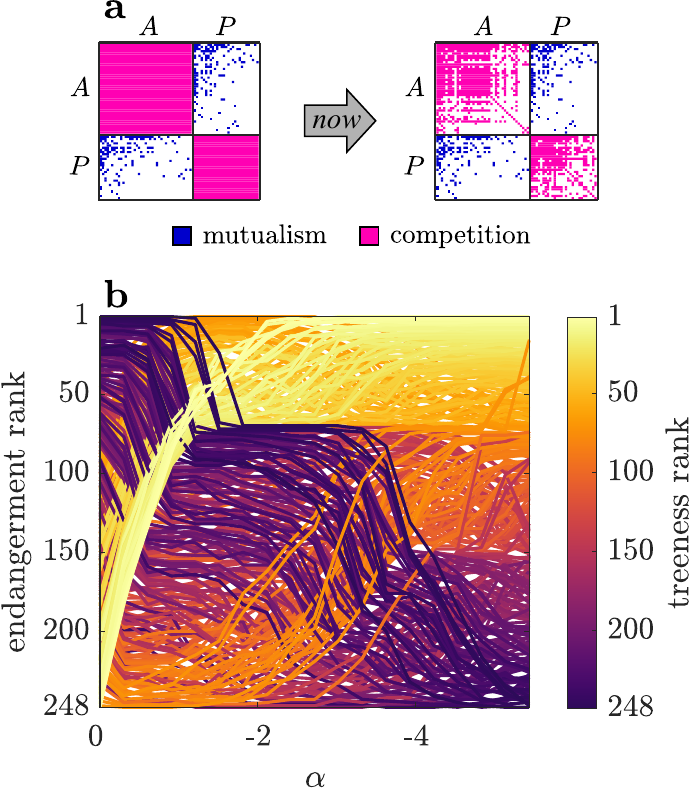}
\caption{\textbf{Evolution of the endangerment ranking with increasing stress level for the case of explicit intra-guild competition topologies.} (a) So far we assumed that species competed with all other species from the same guild in a uniform manner (adjacency matrix on the left side). Now we consider the case of an explicit competition topology which is based on the assumption that the resources for which species compete are their mutualistic partners (adjacency matrix on the right side). (b) Development of endangerment ranks as the intrinsic growth rate $\alpha$ is decreased for the mutualistic model with an explicit competition topology. Color coding of the different species corresponds to treeness rank based on the EV2 of the respective node.}
\label{fig_twocomps_explicit}
\end{figure}

Following the same procedure as before (see Sec. \ref{vary_alpha}) and using the same exemplary plant-pollinator network (see Fig. \ref{fig_2nets}), we check how the endangerment of species within the multilayer setup changes depending on the intrinsic growth rate $\alpha$ (Fig. \ref{fig_twocomps_explicit}(b)). It should again be noted that for the most negative $\alpha$ (right side of Fig. \ref{fig_twocomps_explicit}(b)), the system is close to a bifurcation point beyond which the stability of the desired state would no longer be maintained.

It is striking that, under rather stress-free conditions (high $\alpha$), the obtained endangerment ranking (Fig. \ref{fig_twocomps_explicit}(b)) is nearly reversed when compared to the mean-field competition approach (Fig. \ref{fig_twocomps}(a)). As external stress levels increase, first, the endangerment ranks of the most specialized and most peripheral species of the mutualistic layer drastically increase (bright/yellow colored nodes with high treeness ranks take the upper endangerment ranks until $\alpha \approx -2$), followed by the species with intermediate treeness ranks (orange nodes take the medium endangerment ranks until $\alpha \approx -4$). Ultimately, for high stress levels (small $\alpha$), the endangerment ranking converges towards the ranking which has been obtained for the mean-field competition approach (compare right side of Fig. \ref{fig_twocomps_explicit}(b) with Fig. \ref{fig_twocomps}(a)).

\subsection{Composite Centrality Index}

\begin{figure}[ht]
\centering
\includegraphics[width=0.9\linewidth]{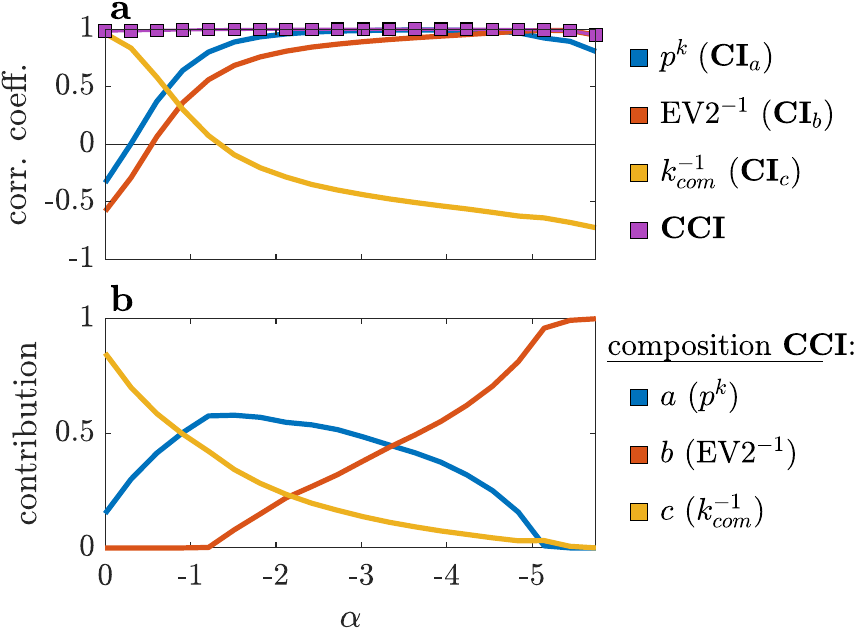}
\caption{\textbf{Topological determinants of endangerment in the case of explicit intra-guild competition.} (a) Linear correlation between the extinction probability $\Omega$ and $p^k$ (blue line), EV2 (orange line), the degree within the competition network $k_{com}$ (yellow line) and the best composite centrality index $\mathbf{CCI}$ (purple squares) which is obtained as the weighted sum of the normalized forms of $p^k$, EV2$^{-1}$ and $k^{-1}_{com}$. (b) Best composition of the $\mathbf{CCI}$ with $a$ describing the contribution of $p^k$, $b$ the contribution of EV2$^{-1}$ and $c$ the contribution of $k^{-1}_{com}$.}
\label{fig_strange_zwei}
\end{figure}

The evolution of the endangerment ranking (Fig. \ref{fig_twocomps_explicit}(b)) indicates that intra-guild competition has a great impact on the endangerment of species when overall conditions are rather favorable (high $\alpha$) but loses in significance as conditions become harsher (smaller $\alpha$) -- in comparison to the impact of inter-guild mutualism (see also \cite{lever2014sudden}). This impression is backed up by the correlation between the endangerment $\Omega$ and a species' degree within its competitive layer $k_{com}$ (Fig. \ref{fig_strange_zwei}(a)), which is high only for high $\alpha$ and decreases fast as conditions become harsher. In return, the fit between $\Omega$ and the two centrality measures referring to the mutualistic layer ($p^{k}$ and EV2) becomes better as the stress level increases (Fig. \ref{fig_strange_zwei}(a)). For the most negative $\alpha$, the best fit with the endangerment is again obtained for the treeness described by the EV2.

We again test whether a composite centrality index $\mathbf{CCI}$ can provide a good fit for all $\alpha$. To cover the impact of the competition layer, we expand the index presented in Eq. (\ref{eq_CCI_one}) by a third normalized centrality index $\mathbf{CI}_c$ and thus obtain
\begin{equation}
    \mathbf{CCI} \; = \; a \, \mathbf{CI}_a \, + \, b \, \mathbf{CI}_b \, + \, c \, \mathbf{CI}_c \;\; ,
\end{equation}
where now $a+b+c=1$ and $a,b,c \ge 0 $. For the third component $\mathbf{CI}_c$, we use the inverse of the degree $k^{-1}_{com}$ within the competitive layer, i.e., $1$ for a pollinator with one competitive link, $1/2$ for a pollinator with two links and so on. We normalize the corresponding vector to a length of $1$, so that $\mathbf{CCI}$ is obtained as the weighted sum of three vectors, each with a length of $1$. Again, we check for each value of $\alpha$ which combination of $a$, $b$ and $c$ provides the highest Pearson's linear correlation coefficient with the extinction probabilities $\Omega$ (Fig. \ref{fig_strange_zwei}(b)). The best $\mathbf{CCI}$ provides a good fit with the endangerment for all $\alpha$ (Fig. \ref{fig_strange_zwei}(a)).

Following the best composition of $\mathbf{CCI}$ with changing $\alpha$ (Fig. \ref{fig_strange_zwei}(b)), we see how the mutualistic layer replaces the competitive layer as the dominant descriptor of pollinator endangerment. For very high $\alpha$ ($\alpha \approx 0$), the degree within the competition matrix $k_{com}$ is the main driver of the endangerment, while the mutualistic degree only provides a minor contribution. Accordingly, species in tree-like substructures of the mutualistic network are least endangered (see Fig. \ref{fig_twocomps_explicit}(b)) -- they have few mutualistic partners but also few competitors. On the contrary, species in the core of the mutualistic network are most endangered as the high number of mutualistic partners cannot compensate the impact of having many competitors. With the decrease of $\alpha$, the impact of $k_{com}$ decreases continuously (Fig. \ref{fig_strange_zwei}(b)). At first, the contribution of the number of mutualistic partners ($p^{k}$) increases in return and becomes the main descriptor of the endangerment for intermediate values of $\alpha$. As a result, the endangerment of specialists increases dramatically and they replace species in the mutualistic core as the most endangered species (see Fig. \ref{fig_twocomps_explicit}(b)). As conditions continue to become harsher, the EV2 becomes a factor (at $\alpha \approx 1.2$). With the increasing impact of the EV2 (Fig. \ref{fig_strange_zwei}(b)), species in trees establish as the most endangered ones (see Fig. \ref{fig_twocomps_explicit}(b)). With the EV2 replacing the mutualistic degree more and more as the main descriptor of the endangerment (Fig. \ref{fig_strange_zwei}(b)), the endangerment more and more resembles the one obtained for the mean-field competition approach -- which means that species in the core are least and species in trees most endangered (see Fig. \ref{fig_twocomps_explicit}(b)). 

\section{Discussion and conclusion} \label{discussion}

The central premise of this work is that we consider mutualistic plant-pollinator networks as dynamical systems exhibiting multistability. This implies that a shock perturbation (i.e., an instantaneous change of system state variables) can induce a shift from a system's current basin of attraction into an alternative basin -- a mechanism also referred to as shock-induced tipping \cite{halekotte2020minimal, vanselow2022evolutionary}. In the mutualistic systems considered, such an event is always accompanied by the loss of some plant and/or pollinator species. To examine the endangerment of pollinators, we calculated an extinction probability for each species based on a set of random shock perturbations which we applied to different realistic network topologies (see Table \ref{table1}). We then ranked the pollinator species within one network from most to least endangered and compared this endangerment ranking to ecologically meaningful centrality metrics obtained from different network-theoretic ranking algorithms. We found that the endangerment of a pollinator species is strongly linked to its degree and its position within the core-periphery structure of its mutualistic network, with the most endangered species being specialists (nodes with low degree) in the outer periphery (outer $k$-shells). Particularly well established instances of such peripheral areas are tree-shaped structures of the network which stem from links between nodes/species in the $1$-shell -- which prompted us to summarize our findings in the deliberately ambiguous slogan \textit{keep the bees off the trees}.

The particular significance of a pollinator species' degree of specialization for its endangerment has already been highlighted in earlier empirical studies \cite{burkle2013plant, aizen2012specialization}. Furthermore, recent theoretical work demonstrated that the positioning within the core-periphery structure of a mutualistic network can be an important factor for a species' endangerment \cite{morone2019k}. Our work affirms such findings but adds a new perspective to the existing theoretical analyses. So far, studies which involved a dynamic description of a plant-pollinator network usually examined the system's response to gradual environmental degradation \cite{morone2019k, lever2014sudden, dakos2014critical, jiang2018predicting, lever2020foreseeing, aparicio2021structure}. Our work complements such studies by capturing another section of the spectrum of potential stressors. The abrupt and large shock perturbations, which we consider, can be interpreted as non-specific extreme events -- a class of disturbances which plants and pollinators are likely to experience (e.g., due to wildfires, extreme rainfall or sudden pesticide exposure). Another difference to most former studies is that, instead of highlighting the point of collapse (i.e., the loss of most or a significant amount of species) or the precursory signs thereof \cite{aparicio2021structure, bascompte2023resilience}, we consider all possible outcomes of a perturbation (i.e., all coexisting attractors), including major as well as minor extinction events.

It is this approach which allowed us to distinguish all pollinator species within a mutualistic network based on their individual chance of getting extinct. The derived endangerment ranking then enabled us to examine how topological traits of nodes affect the endangerment of corresponding species. In this regard, the finding that topology-based metrics which reflect the "treeness" of a node -- in particular, the inverse eigenvector centrality type II (EV2) and the inverse neighborhood centrality based on the $k$-shell index ($C_N$) -- are well suited to capture the overall distribution of vulnerabilities is instructive in multiple ways. Firstly and now obviously, it highlights the potential endangerment of species engaging in mutually specialized interactions (species in trees). This might not only concern the endangerment of currently present pollinator species but could also give a hint at why corresponding motifs, like tree-like substructures, are rather rare in plant-pollinator networks. Another aspect refers to the usual use of the applied ranking algorithms. Centrality metrics, like the eigenvector and neighborhood centrality, have been developed to identify the most important nodes within networks. Accordingly, losing the most endangered species, corresponding to the nodes with the lowest centrality (or the highest inverse centrality), is unlikely to profoundly affect the overall network integrity. While this is certainly no bad news, it also implies that many species could be lost long before a mutualistic network is anywhere close to a system collapse -- i.e., many species could be extirpated without warning.

The main focus of this work was examining the impact of network topology on species endangerment. We therefore chose a simple parametrization scheme which ensures that species solely differ on account of their position within the mutualistic network. This also meant neglecting some phenomena which occur in real plant-pollinator networks, like phenological dynamics which affect the topology of mutualistic networks over the course of a season \cite{kaiser2010robustness, miele2020core} or differences in the level of reliance on mutualistic partner species \cite{vieira2015simple, vanbergen2017network, bluthgen2010network} (e.g., some pollinators also feed on resources outside the mutualistic network). Due to the simplistic approach, it should be obvious that our findings cannot be directly applied to assess the endangerment of real world pollinator species. In fact, under certain conditions, the mutual specialization might have had benefits for the involved species which enabled the evolution and persistence of this phenomenon. To shed some light on potential benefits of being located in the periphery, we considered an explicit topology for the intra-guild competition (Sec. \ref{explicit_comp}) which is based on the assumption that the resources for which species compete are their mutualistic partners \cite{gracia2018joint}. We found that being peripheral can be beneficial when conditions are good, since it allows pollinators to avoid competition for resources. However, we also saw that, under comparatively harsh conditions, pollinators in tree-like structures are again the ones being most endangered -- i.e., with increasing stress, mutualism rather than competition becomes the decisive interaction for a species' survival (in agreement with the \textit{stress-gradient hypothesis} \cite{bertness1994positive, he2013global}). Accordingly and in the light of the ongoing degradation of the living conditions of pollinator species in the real world \cite{soroye2020climate, siviter2021agrochemicals, dicks2021global}, our catchphrase \textit{keep the bees off the trees} might be of significance now more than ever.

In the real world, the multilayer nature of plant-pollinator systems goes way beyond the two layers which we considered in this work. For instance, bees require both nectar and pollen which they often receive from different flowers \cite{westrich2018wildbienen}. Accordingly, a degree greater than 1 does not necessarily mean that a species is not specialized in one way or the other. The same holds true for competition since species do not only compete for mutualistic partners but also for nesting sites (pollinators), and nutrients and space (plants). Overall, we believe that the interplay as well as potential trade-offs between different network layers are aspects which deserve further investigation in future studies. The approach which we presented in this work -- i.e., creating endangerment rankings and comparing them to network-based metrics -- can be an instructive tool for such analyses. In particular, it is one which can be easily adopted for other, potentially more complex, disturbance scenarios like periodically recurring disturbances \cite{meyer2018quantifying}, disturbances with an explicit temporal structure \cite{vanselow2019very} or combinations of gradual environmental degradation and shock perturbations \cite{harris2018biological}. But for now, in reference to single shock perturbations and to our simple system setup, we stick to our initial recommendation to \textit{keep the bees off the trees} (trees in the network theoretical sense).

\begin{acknowledgments}
The simulations were performed at the HPC Cluster CARL, located at the University of Oldenburg (Germany) and funded by the DFG through its Major Research Instrumentation Programme (INST 184/157-1 FUGG) and the Ministry of Science and Culture (MWK) of the Lower Saxony State.
\end{acknowledgments}

\appendix

\section{Plant-pollinator networks}\label{appendix_weboflife}
The topologies of all studied plant-pollinator networks have been taken from the Web of Life database (\url{www.web-of-life.es}). In order to attain a set of comparable, connected networks of sufficient size, we processed the original data as follows: (1) All networks are considered as being unweighted. (2) From each dataset, only the largest connected component is taken while all other components are omitted. (3) Only those networks which hold more than 100 topologically unique pollinator nodes $\tilde{N}_A$ (i.e., nodes which have a unique set of neighbors) are considered -- so that we end up with a set of 11 networks (see Table \ref{table1}). The reason for the last selection criterion is that we do not consider all but only the distinguishable nodes for the endangerment and centrality rankings (see Sec. \ref{sec_cen_vs_end}).

\begin{table}[htb]
\caption{\label{table1}
Overview of the dataset of applied plant-pollinator ($P$-$A$) networks. From each dataset only the largest connected component is considered. For the ranking only the unique pollinator nodes ($\tilde{N}_A$) are taken into account.}
\begin{ruledtabular}
\begin{tabular}{ccccc}
\textrm{ID\footnote{In the web of life database (\url{www.web-of-life.es}).}}&
\textrm{source\footnote{Original study in which the network topology was established.}}&
\textrm{\#links\footnote{\# mutualistic links in the largest connected component.}}&
\textrm{~$N_P$ / $N_A$\footnote{\# plants and pollinators in the largest connected component.}}&
\textrm{$\tilde{N}_A$\footnote{\# unique pollinator species (considered in the ranking).}}\\
\colrule
M\_PL\_005 & \cite{clements1923experimental} & 918 & ~91 / 270  & 170\\
M\_PL\_015 & \cite{petanidou1991pollination} & 2930 & 130 / 663 & 476\\
M\_PL\_021 & \cite{kato1990insect} & 1192 & ~90 / 676  & 206\\
M\_PL\_044 & \cite{kato2000anthophilous} & 1121 & 107 / 605 & 248\\
M\_PL\_048 & \cite{dupont2009ecological} & 671 & ~30 / 236  & 132\\
M\_PL\_049 & \cite{bek2006pollination} & 590 & ~37 / 225  & 118\\
M\_PL\_053 & \cite{yamazaki2003flowering} & 567 & ~92 / 272  & 139\\
M\_PL\_054 & \cite{kakutani1990insect} & 763 & 106 / 308 & 167\\
M\_PL\_055 & \cite{kato1996flowering} & 427 & ~61 / 192  & 117\\
M\_PL\_056 & \cite{kato1993flowering} & 871 & ~91 / 365  & 188\\
M\_PL\_057 & \cite{inoue1990insect} & 1920 & 114 / 883 & 319\\
\end{tabular}
\end{ruledtabular}
\end{table}

\section{Dynamical model set-up}\label{appendix_parameters}
For the choice of parameters in the model of plant-pollinator networks (see Sec. \ref{basics_model}), two considerations were decisive. The first refers to the main goal of this work, which is to link topological traits of network nodes to the endangerment of corresponding species. To ensure that differences in the endangerment (given by $\Omega$) actually reflect differences in the topological traits of the respective nodes, we choose a parametrization which makes sure that species solely differ on account of their position within the mutualistic network. This is achieved by using the same set of constant parameters for each species $i$ (e.g., $\alpha_i=\alpha\;\forall\;i$). The second important consideration concerns the existence (and attractiveness) of the desired state $\mathbf{X_0}$ in which all species coexist. Since $\mathbf{X_0}$ serves as the ground state for the applied perturbation scheme, which means that the system resides in this state prior to a shock perturbation (see Sec. \ref{basics_perturb}), the desired state $\mathbf{X_0}$ should be locally attractive (linearly stable) for each of the examined mutualistic systems (see Table \ref{table1}). We choose the parameters accordingly, with the standard setting being $\alpha = 1.0$, $\beta_{ii}=1.0$, $\gamma_0 = 10.0$, $h=0.1$ and $\zeta=0.5$ (with $\alpha$ being the only parameter which is varied within this work, in Sec. \ref{results_robustness} and Sec. \ref{explicit_comp}). Aside from these parameters, the parametrization of the interspecific competition and the parametrization of the Allee effect deserve special consideration (see below). 

\textbf{Competition I -- standard setup:} Through most of this work (Sec. \ref{results_ranking} and \ref{results_robustness}) we stick to the commonly applied \textit{competition of mean-field type} \cite{bastolla2009architecture}. In this approach, it is assumed that each species within one guild competes with every other species in the same guild in a uniform manner, i.e., every pollinator competes with every other pollinator and every plant with every other plant (see left side of Fig. \ref{fig_twocomps_explicit}(a)). To obtain a stable desired state $\mathbf{X_0}$ for networks of different size (see Table \ref{table1}), we assume that the strength of competition between two species is mitigated in accordance with the total number of competitors within the network -- i.e., $\beta^P_{il}=\beta_0/(N_{P}-1)$ for $i \neq l$ ($\beta^A_{jo}=\beta_0/(N_{A}-1)$ for $j \neq o$), where $N_P$ and $N_A$ are the total number of plant and pollinator species in the network and $\beta_0=1.5$.

\textbf{Competition II -- multilayer setup:}  In Sec. \ref{explicit_comp}, we consider an alternative competition topology (see right side of Fig. \ref{fig_twocomps_explicit}(a)) which we derive from the network of mutualistic interactions in the following manner: A competitive link is drawn between two distinct species ($i \neq l$) from the same guild if they share at least one mutual partner ($\beta_{il}>0$), otherwise no link is drawn ($\beta_{il}=0$). For the sake of simplicity, we assume that the strength of competition is the same for every pair of competing species -- i.e., either $\beta_{il}=0$ or $\beta_{il}=0.001$ for $i \neq l$ (applies to both plant competition $\beta^P$ and pollinator competition $\beta^A$).

\textbf{The Allee effect:} Species with lower abundance $P_i^*$ ($A_j^*$) in the desired state $\mathbf{X_0}$ generally have a higher chance of obtaining a low absolute abundance $P_i(t=0)$ ($A_j(t=0)$) after the shock perturbation (see Sec. \ref{basics_perturb}). Since the Allee effect $q_i$ depends on the species abundance $P_i$ (see Eq. (\ref{Eq_allee})), species with low $P_i^*$ would be disproportionately penalized if we would use the same $\theta_i$ for all species $i$. In order to avoid this, we derive an individual $\theta_i$ for every species $i$ which depends on the species' abundance $P_i^*$ in the desired state. To achieve this, we divide the parametrization into two stages. In the first stage, we set up a provisional system without an Allee effect (corresponds to setting $q_i=1$ for all $i$) and numerically determine the corresponding provisional desired state $\mathbf{\tilde{X}_0}$. In the second stage, we obtain $\theta_i$ as  
\begin{equation} \label{Eq_theta}
    \theta_i = \frac{-0.1 \tilde{P}_i^*}{\ln(0.5)} \; ,
\end{equation} where $\tilde{P}^*_i$ is the abundance of species $i$ in the provisional desired state $\mathbf{\tilde{X}_0}$. Since $\tilde{P}^*_i \approx P^*_i$, the Allee effect $q_i$ now depends on a species' relative abundance $P_i/P^*_i$ (insert Eq. (\ref{Eq_theta}) into Eq. (\ref{Eq_allee})). Due to the choice of $\theta_i$, $q_i$ takes a value of $0.5$ when species $i$ is at $10\%$ of its abundance $P^*_i$ in the desired state.

\section{Allee effect induces degree-dependence}\label{appendix_allee}

In the following, we demonstrate that if we assume that mutualism is obligatory ($f_i<0$) and that any species can go extinct if perturbed strong enough (Allee effect $q_i$), a dependence of the endangerment on the degree is inherent in the model of mutualism (Eq. (\ref{Eq_mutualism})). To this end, we reduce the model to the necessary ingredients allowing for an extinction threshold and obligatory mutualism. As a first step, we neglect the interspecific competition, $g_i=0$. Moreover, we assume that a species obtains the full mutualistic benefit as long as it has any partner species left. This effectively reduces the mutualistic benefit $m_i$ to an ON-OFF function which reads
\begin{equation} \label{eq_onoff}
    m_i \, = \, \begin{cases}
\text{const} &  \text{if any} \;\, \gamma^P_{ij} A_j > 0\\
0 & \text{else} \; .
\end{cases}
\end{equation}
It should be noted that this form of $m_i$ can be obtained by assuming an extremely efficient mutualism, $\gamma_0 \to \infty$, in which case, the constant in Eq. (\ref{eq_onoff}) is $h^{-1}$.

Under the assumption that we only consider connected networks, the extinction of a species in this simplified model can initially be induced only due to its own density falling below the Allee threshold. Further extinctions can occur if a species loses all its partners due to such initial extinction events. Accordingly, the extinction process following a single large shock perturbation can be reformulated by two simple probabilistic rules which denote primary and secondary extinctions.
\begin{itemize}
    \item[(1)] Primary Extinctions: Each species has a probability $p_i$ of going extinct due to the shock.
    \item[(2)] Secondary Extinctions: After the initial shock, species which remain without any mutualistic partner are lost as well. The probability for this to occur is the product of the primary extinction probabilities $p_j$ of the neighbors $\Gamma_i$ of species $i$.
\end{itemize} Accordingly, the probability of species $i$ to go extinct due a single shock perturbation can be denoted as 
\begin{equation}
 \Omega_i \, = \,  p_i + (1-p_i) \prod_{j \in \Gamma_i}p_j \; .
\end{equation}

For the sake of simplification and in accordance with the parametrization of the Allee effect (see Appendix \ref{appendix_parameters}), we assume that the probability of falling below the primary extinction threshold is the same for all species and thus $p_i=p \; \forall \; i$. This allows us to derive an analytic solution for the extinction probability for each species which solely differs on account of a species' degree $k_i$
\begin{equation} \label{endangerment_simple}
    \Omega_i \, = \,  p + (1-p)p^{k_i} \; .
\end{equation}
Accordingly, in this simple model, each species holds a certain basal-endangerment $p$ which yields the minimum for the $\Omega_i$ of any species and an additional endangerment term whose contribution to $\Omega_i$ decreases with increasing degree $k_i$ according to the power law $p^{k_i}$.

\begin{figure}[ht]
\centering
\includegraphics[width=0.8\linewidth]{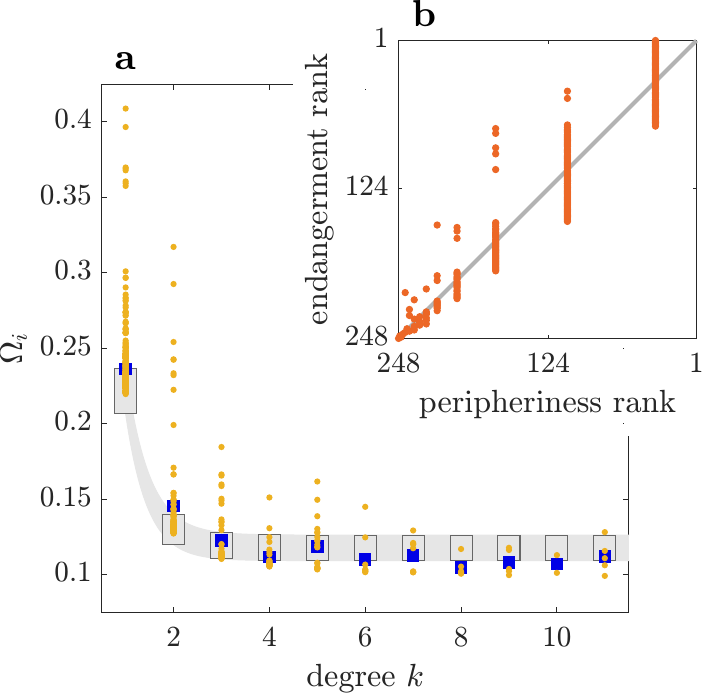}
\caption{\textbf{Comparison between the endangerment in the complete and the strongly simplified model.} (a) Dependence of the extinction probability $\Omega_i$ on the degree. Yellow dots correspond to $\Omega_i$ of species in the dynamic model of the exemplary mutualistic system (M\_PL\_044 in Table \ref{table1}). Blue squares represent the median of $\Omega$ for all species with the same degree. Gray boxes correspond to the $\Omega_i$ provided by the simplified model (Eq. (\ref{endangerment_simple})) where the upper edge is determined by adaptation to the specialists ($k=1$) and the lower bound by adaptation to the generalists ($k>7$). (b) Rank correlation between peripheriness (inverse centrality) measured by degree and endangerment measured by $\Omega$.}
\label{fig_specialists}
\end{figure}

In the following, we test whether the extinction probabilities $\Omega$ calculated for the complete dynamical model generally follow the dependency described by Eq. (\ref{endangerment_simple}). To this end, we need to derive a basal-endangerment $p$ for the simplified model (Eq. (\ref{endangerment_simple})) based on the extinction probabilities obtained for the complete model. We determine two instances of $p$, representing an upper limit $p_{up}$ and a lower limit $p_{low}$ (upper and lower edge of gray boxes in Fig. \ref{fig_specialists}(a)). The upper limit $p_{up}$ is obtained by solving Eq. (\ref{endangerment_simple}) for the median of the extinction probabilities $\Omega$ of all species with a degree $k=1$ in the exemplary network, while $p_{low}$ is set to the median of the extinction probabilities $\Omega$ of all species with a degree $k>7$, as $p^k \to 0$ for $p \ll 1$ and $k \gg 1$.

The comparison shows that the endangerment distribution obtained from the complete dynamical model generally follows the dependency described by Eq. (\ref{endangerment_simple}) (Fig. \ref{fig_specialists}(a)). Accordingly, the degree is an important driver of a species' endangerment. However, we observe that species with the same degree strongly differ in their endangerment and that the simple approximation is not able to capture the exceptional endangerment of some species (especially in the specialist class). Accordingly, the degree is an important but not the only driver of a species' endangerment.

We furthermore find that, if adapted to the endangerment of generalists ($p_{low}$), the simplified model underestimates the endangerment of specialists. This indicates that the simplification might not capture all aspects which cause the particular endangerment of specialists -- for instance, the simplified model does not take into account that shock perturbations affect the mutualistic term $m_i$ (Eq. (\ref{eq_m}), but see Appendix \ref{appendix_irwinhall}).

\section{Mutualistic benefit after a shock}\label{appendix_irwinhall}
The relation between the degree of specialization and the endangerment of a species derived in Appendix \ref{appendix_allee} was based on the assumption that mutualistic benefits were fully present (saturated) as long as any partner species was left. This does of course neglect an integral element of the mutualistic system, which is the term describing the actual mutualistic benefit $m_i$ a species obtains from the interaction with its partners (Eq. (\ref{eq_m})).

The mutualistic benefit $m_i$ depends on the number and abundance of partner species. Accordingly, since the abundances of species constitute the state variables of the system, the mutualistic benefit is a dynamical time-dependent quantity, $m_i(\mathbf{A}(t))$. However, for the sake of simplicity, we consider $m_i$ at one particular point in time, $t=0$, which is the time at which the shock perturbation has just hit the system. In other words, we simply examine how the perturbation shapes the mutualistic benefit $m_i$. At $t=0$, the abundance of each species can be considered as a random variable drawn from a uniform distribution in the interval $[0, N^*_j]$, where $N^*_j$ is the abundance of species $j$ in the desired or pre-disturbance state (see Sec. \ref{basics_perturb}). Assuming for now that $N^*_j$ is the same for all species ($N^*_j=N^*$), we can derive a probability distribution for the mutualistic benefit $m_i$ using a transformation of the Irwin-Hall distribution \cite{johnson1995continuous}. The resulting probability density function is
\begin{eqnarray} \label{eq_prob}
    f_y = \frac{1}{(k-1)!} && \sum_{j=0}^{\left\lfloor\frac{y}{\gamma N^*(1-hy)}\right\rfloor} (-1)^j  \binom{k}{j} \frac{1}{\gamma N^*(1-hy)^2} \nonumber \\ &\cdot& \left( \frac{y}{\gamma N^*-\gamma N^*hy}-j \right)^{k-1} ~ ,
\end{eqnarray}
with $\gamma = \gamma_i = \gamma_0/k_i^{\zeta}$ and $\lfloor \cdot \rfloor$ being the floor function.

\begin{figure}[htb]
\centering
\includegraphics[width=1.0\linewidth]{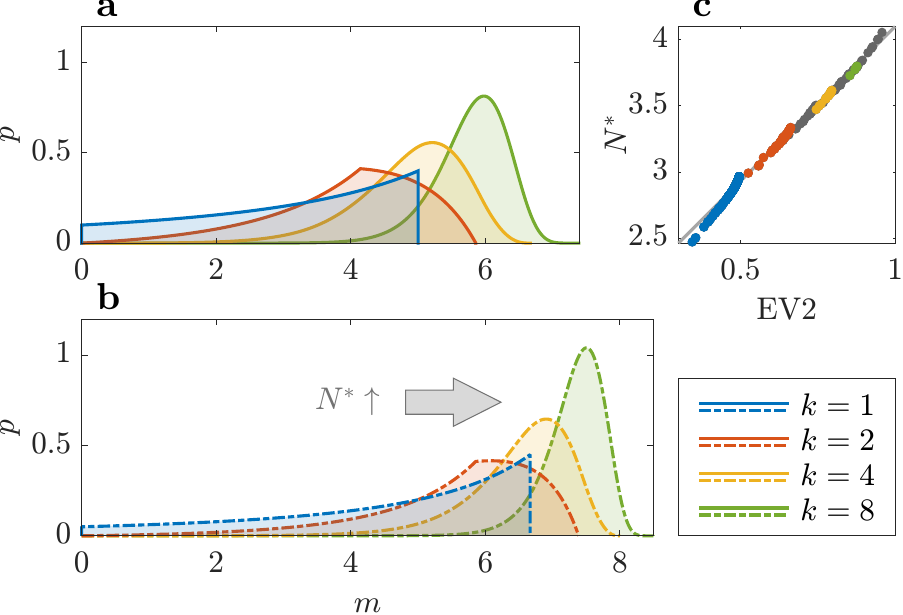}
\caption{\textbf{Probabilistic description of the mutualistic term $m_i$ at $t=0$.} (a,b) Probability distribution of the mutualistic benefit for different degrees $k$ for $N^*=1.0$ (a) and $N^*=2.0$ (b). (c) Correlation between the EV2 and the abundance of species $N^*$ in the desired state in the exemplary mutualistic system (M\_PL\_044 in Table \ref{table1}).}
\label{fig_prob}
\end{figure}

Admittedly, Eq. (\ref{eq_prob}) is a little difficult to read. Therefore, to understand the impact of the degree $k$ and the abundance of partners $N^*$ on the mutualistic benefit $m_i$ which remains after a perturbation, the inspection of exemplary probability distributions for specific $k$ and $N^*$ is instructive (Fig. \ref{fig_prob}(ab)). Regarding the impact of the remaining mutualistic benefit on the endangerment of a species, it can be assumed that very low $m_i(t=0)$ are particularly dangerous since they can cause an overall negative growth rate (Eq. (\ref{Eq_mutualism})) at the time of the shock perturbation ($t=0$).

Regarding the impact of the degree $k$, we note that the degree affects both the position and the shape of the probability distribution of the mutualistic benefit $m_i(t=0)$ (Fig. \ref{fig_prob}(a)). Due to the specific shape of the distribution for low degrees, especially for $k=1$, very low $m_i(t=0)$ are way more probable for specialists than for less specialized species ($k>1$). The way in which the mutualistic benefit $m_i$ is affected by the shock perturbation thus reveals another aspect amplifying the particular endangerment of specialists.

In contrast to the degree, the abundance of partners $N^*$ mainly affects the position of the probability distribution of $m_i(t=0)$ -- an increase of $N^*$ leads to a distortion of the probability distribution towards larger mutualistic benefits $m_i$, while the specific shape of the curve is basically maintained (see change from Fig. \ref{fig_prob}(a) to Fig. \ref{fig_prob}(b)). This means that the chance of receiving a very low $m_i$ right after a perturbation decreases significantly for species whose partners have a high abundance in the undisturbed system state. However, the abundance $N^*$ is not the same for all species but relies on the species' position within the network. In fact, we find a strong correlation between the abundance $N_i^*$ and the centrality of the corresponding node, a relation which is well captured by the EV2 (Fig. \ref{fig_prob}(c)). Accordingly, species whose partners show a low EV2 score -- corresponding to peripheral nodes -- have a higher chance of receiving a low mutualistic benefit after the shock perturbation than species which are linked to species in the core of the network (high EV2).

% \bibliographystyle{ieeetr}
% \bibliography{references}
\end{document}